601.01

# Detecting Exoplanets in the Presence of Exozodiacal Dust Profiles

### Charley Noecker (BATC), Marc Kuchner (GSFC)

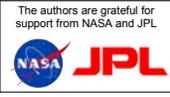 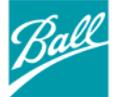 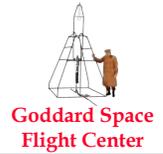

The authors are grateful for support from NASA and JPL

Goddard Space Flight Center

**Abstract:**
For exoplanet direct detection mission concepts such as Terrestrial Planet Finder or Exoplanet Probe, light from the exozodiacal dust tends to obscure any exoplanets present in the image. Data analysis methods to identify point sources against this background have been very simple, traditionally with the simplifying assumption that the exozodi is uniformly distributed, just as our local zodiacal background is uniform over several-arcsec scales. However, the typical size of an exozodi cloud is expected to be comparable to the typical exoplanet orbital radii, or at least those of greatest interest — the "habitable zone" range from 0.7-1.5 AU.

When a direct detection instrument is reduced in size for cost reasons, the point spread function (PSF) becomes broader, making it more difficult to distinguish a point source from a "blob" of exozodi light. In this case, the shot-noise limited integration time may not be enough; instead we may need an elevated signal-to-noise ratio and/or later measurements to resolve ambiguities in the image data, identify a point source with a calculable and high confidence level, and isolate the exozodi and exoplanet contributions to the observed light profile. We will examine some typical profiles and a few methods of analyzing image data, with the goal of structuring an approach to this data analysis problem.

## The Search for Extrasolar Planets

- The Exoplanet Exploration Program (ExEP, exep.jpl.nasa.gov) is leading NASA's effort to find and characterize exoplanets, including by direct imaging methods.
- For direct imaging detection and spectroscopic characterization of Earth-size exoplanets at Earth-like radii, several mission concepts are being considered,
  – Terrestrial Planet Finder (flagship)
  – Exoplanet Probe

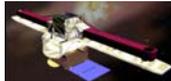
FKSI

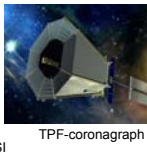
TPF-coronagraph

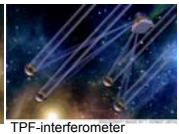
TPF-interferometer

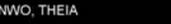
NWO, THEIA

PECO
ACCESS
EPIC

## The Challenge of Exozodiacal Light

We expect every target star to have zodiacal dust clouds, of unknown brightness
**Exozodi increases integration times**
➔ DC background for the planet signal
➔ increases shot noise and integration times

**Exozodi profile features can mimic small exoplanets**
- Exozodi profile traditionally assumed <u>symmetric</u> or even <u>uniform</u>
  – Neglects difficulties in fitting exozodi and isolating the planet
- Asymmetry in the image can be caused by
  – Dynamics of the planetary system (giant planets ➔ pericenter shift)
  – Occulter decenter and asymmetry
  – Requires a more general exozodi profile model for planet detection analysis
- Broad PSF (small telescope) leaves ambiguity between point sources and exozodi features

**Terrestrial planet detection requires fitting a fully general exozodi model**
- From panchromatic camera or integral field spectrograph data
- Fit exozodi shape and brightness, coronagraph suppression, and exoplanet position
- Correct for expected exozodi profile features, based on measurements
- Uncertainty in <u>exozodi profile</u> increases uncertainties in <u>exoplanet position</u>, <u>brightness</u>, and <u>spectrum</u>

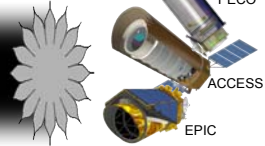

Exozodi plus Earth
Occulter FWHM 120 mas
PSF FWHM = 15×30 mas

Exozodi with 0.1AU pericenter offset

## Causes of exozodi asymmetry

**Planetary dynamics causes <u>real</u> exozodi asymmetry**
- Gravitational influence of giant planet(s) causes exozodi pericenter offset
- Offset is locked to the giant planet's axis of orbital symmetry (periastron-apastron)
- Precise knowledge of giant planet orbits might allow estimates of the asymmetry

**Occulter decentering causes <u>apparent</u> exozodi asymmetry**
- Exozodi brightness profile is sharply peaked near center
- Decentered occulter (internal/external) asymmetrically blocks the exozodi cloud
- Precise knowledge of the offset vs. the star would allow estimates of the apparent asymmetry

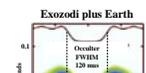
Occulter transmission / Exozodi flux — Exozodi pulled off-center by giant planet

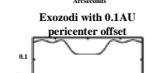
Occulter transmission / Exozodi flux — Occulter aligned off-center

## Measurement limitations

**Broad PSF (small telescope) exacerbates the ambiguity problem**
- Makes point sources (exoplanets) difficult to distinguish from exozodi features
- Features of exozodi clouds are ~0.1-1 AU in size;
- Coronagraph with IWA = 2λ/D has PSF FWHM ≈ 0.5 AU
- A broad PSF washes out details and spreads point sources

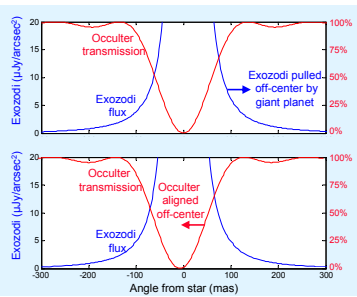
1.5m   2.4m   4m   10m
Cash et al., Proc. SPIE **7010**, p. 70101Q (Marseille, 2008)

## R-φ components of exozodi profiles

Exozodi at 10pc, 10× brighter than Solar zodi, no Sun and no planets, λ=0.5 μm, 60° inclination

$\log_{10}$ contours, jansky/(2.5mas)$^2$

Mark a circle on the sky at each radius; look for sine and cosine harmonics around that circle

Cosines (x): Even harmonics dominate (➔ symmetric left-right)

Sines (y): Odd harmonics dominate (asymmetric top-bottom)

### Baseline case
**Symmetric exozodi, No planets**
Same zodi seen through a 4m diam Lyot coronagraph, jinc$^2$ mask (circular)
IWA = 3.1 λ/D = 80mas
Linear contours

### With Exoplanet
**Symmetric exozodi, Earth-like planet**
Same zodi with Earth-like planet (on the left side), Seen through the same coronagraph
Linear contours

### Pericenter Offset
**Asymmetric exozodi, No planets**
Same zodi with pericenter offset of 0.0131 AU, Seen through the same coronagraph
Linear contours

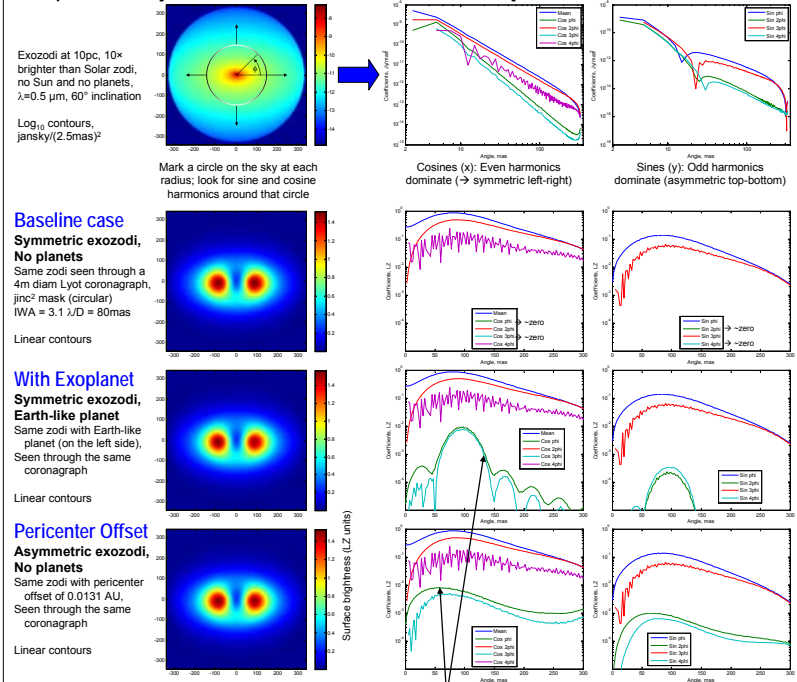

The differences between these 2 curves are < 1% of the average EZ flux
Demands greater measurement sensitivity to distinguish the 2 possibilities

## Modeling uncertainty in exozodi profile

The example pericenter offset is comparable to
- Dynamical effect of our Jupiter on our local zodi
- One Earth signature in the coronagraph
- 2% (TBR) skew asymmetry in internal occulter attenuation profile
- 1.3 mas offset of external occulter from star (0.5 m)
- 1.3 mas error in locating the star on the detector
- 10× smaller pericenter offset for 10× brighter exozodi

- Exozodi modeling could become the dominant uncertainty / ambiguity in finding exoplanets
- Small telescopes (fat PSFs) make it more difficult
- Brighter exozodis make it more difficult

## Possible mitigation strategies

- Increased telescope size to sharpen the PSF
- Increased SNR to aid in distinguishing subtle shape variations
- Multiple revisits to same target star to observe exoplanet orbital motion
- Advanced exozodi models supported by calibration measurements

### Conclusions:

(Exozodi asymmetry) ≈ (Fat PSF) ⊗ (Faint planet)

➔ Faint-planet detection requires careful modeling of exozodi asymmetry
➔ Uncertainty in faint-planet detection can be limited by uncertainty in modeling the exozodi asymmetry